\documentstyle[aps,epsfig]{revtex}

\begin{document}

\widetext 

\title{\bf A model independent approach to non dissipative decoherence}
\author{Rodolfo Bonifacio$^{1}$, Stefano Olivares$^{1}$, 
Paolo Tombesi$^{2}$, and David Vitali$^{2}$}
\address{$^{1}$Dipartimento di Fisica, Universit\`a di Milano, \\
INFN and INFM, Sezione di Milano, via Celoria 16, 20133, Milano, Italy \\
$^{2}$Dipartimento di Matematica e Fisica, Universit\`a di
Camerino, \\
INFM, Unit\`a di Camerino, via Madonna delle Carceri 62032, Camerino, Italy}

\date{\today}

\maketitle

\begin{abstract}
We consider the case when decoherence is due to the fluctuations
of some classical variable or parameter of a system and not to its entanglement
with the environment. Under few and quite general assumptions, we derive a
model-independent formalism for this non-dissipative decoherence,
and we apply it to explain the decoherence observed in some recent
experiments in cavity QED and on trapped ions.
\end{abstract}


\section{Introduction}

Decoherence is the rapid
transformation of a pure linear superposition state into the 
corresponding statistical mixture
\begin{equation}
|\psi\rangle=\alpha |\psi _{1} \rangle +
\beta |\psi _{2} \rangle \Rightarrow \rho_{mix}=
|\alpha|^{2}|\psi_{1}\rangle\langle \psi _{1} |+
|\beta|^{2}|\psi_{2}\rangle \langle \psi _{2} | \; ;
\label{decoh}
\end{equation}
this process does not preserve the purity of the state, that is,
${\rm Tr}\rho_{mix}^{2} < 1$, and therefore it has to be
described in terms of a non-unitary evolution. The most common approach
is the so-called environment-induced decoherence 
\cite{zur,zeh} which is
based on the consideration that it is extremely
difficult to isolate perfectly a system from uncontrollable degrees
of freedom (the ``environment''). The non-unitary evolution 
of the system of interest is obtained by 
considering the interaction with these uncontrolled degrees of
freedom and tracing over them. In this approach, decoherence
is caused by the entanglement of the two states of the superposition
with two approximately orthogonal states of the environment 
$|E_{1}\rangle $ and $|E_{2}\rangle $
\begin{equation}
\left(\alpha | \psi _{1} \rangle +
\beta |\psi _{2} \rangle\right) \otimes |E_{0}\rangle \Rightarrow
|\psi\rangle_{TOT}=
\alpha  |\psi _{1} \rangle \otimes |E_{1}\rangle +
\beta |\psi _{2} \rangle \otimes |E_{2}\rangle \;.
\label{entenv} 
\end{equation}
Tracing over the environment and using $ \langle E_{1}|E_{2}\rangle \simeq 
0 $, one gets
\begin{equation} 
{\rm Tr}_{env}\left\{|\psi\rangle_{TOT}\langle \psi |\right\} = \rho_{mix}
\label{romix} \;,
\end{equation}
where $\rho_{mix} $ is defined in Eq.~(\ref{decoh}).
The environment behaves as a measurement 
apparatus because the states $|E_{i}\rangle $ behave as  ``pointer 
states'' associated with
$|\psi_{i}\rangle $; in this way the environment acquires 
``information'' on the 
system state and therefore decoherence 
is described as an irreversible flow of information from the 
system into the environment \cite{zur}. In this approach, the system
energy is usually not conserved and the interaction with the environment 
also accounts for the irreversible thermalization of the system of 
interest.
However this approach is inevitably {\em model-dependent}, because
one has to assume a model Hamiltonian for the environment and the 
interaction between system and environment. This modelization, and 
therefore any quantitative prediction, becomes
problematic whenever the environmental degrees of freedom 
responsible for decoherence are not easily recognizable.

Decoherence is not always necessarily due to the entanglement with
an environment, but it may be due, as well, to the fluctuations of 
some classical parameter or internal variable of the system. 
This kind of decoherence
is present even in isolated systems, where environment-induced
decoherence has to be neglected. In these cases the system energy 
is conserved, and one has a different form of decoherence,
which we shall call ``non-dissipative decoherence''. 
In such cases, every single experimental run is characterized by 
the usual unitary evolution generated by the system Hamiltonian.
However, definite statistical prediction are obtained only
repeating the experiment many times and this is when decoherence
takes place, because each run corresponds to a different random value
or stochastic realization of the fluctuating classical variable.
The experimental results correspond therefore to an average over these
fluctuations and they will describe in general an effective 
non-unitary evolution.

In this paper we shall present a quite general theory of non-dissipative
decoherence for isolated systems which can be applied for two 
different kinds of fluctuating variables or parameters: the case of a 
random evolution time and the case of a fluctuating Rabi frequency
yielding a fluctuation of the Hamiltonian. In 
both cases one has random phases $e^{-iE_{n}t/\hbar}$
in the energy eigenstates basis that, once
averaged over many experimental runs, lead to the decay of off-diagonal
matrix elements of the density operator, while
leaving the diagonal ones unchanged.

The outline of the paper is as follows. In Section II we shall
derive the theory under general assumptions, following closely
the original derivation presented in \cite{rb1,rb2}. In Section III
we shall apply this theory in order to describe the decoherence
effects observed in two cavity QED experiments performed in Paris, one
describing Rabi oscillations associated with the resonant interaction
between a Rydberg atom and a microwave cavity mode \cite{haro}, 
and the second one 
a Ramsey interferometry experiment using a dispersive 
interaction between the cavity mode and the atom \cite{disp}.
In Section IV we shall apply our approach to 
a Rabi oscillation experiment for trapped ions \cite{wine}, and
Section V is for concluding remarks.

\section{The general formalism}

The formalism describing non-dissipative decoherence of isolated
systems has been derived in \cite{rb1,rb2} by considering the case of a 
system with random evolution time. The evolution time may be random 
because of the finite time needed to prepare the initial state of the 
system, because of the randomness of the detection time, as well as 
many other reasons. For example, in cavity QED experiments, the 
evolution time is the interaction time, which is determined by the 
time of flight of the atoms within the cavity and this time can be 
random due to atomic velocity dispersion.

In these cases, the experimental observations are 
not described by the usual density matrix of the whole system 
$\rho(t)$, but by its
time averaged counterpart \cite{rb1,rb2}
\begin{equation}
\bar{\rho}(t)= \int_{0}^{\infty} dt' P(t,t') \rho(t') \;,
\label{robar} 
\end{equation}
where $\rho(t') = \exp\{-iLt'\}\rho(0)$ is the usual unitarily evolved
density operator from the initial state and $L\ldots =[H,\ldots]/
\hbar$. Therefore $t'$ denotes the random evolution time, while
$t$ is a parameter describing the usual ``clock'' time.
Using Eq.~(\ref{robar}), one can write
\begin{equation}
\bar{\rho}(t)= V(t) \rho(0) \;,
\label{robar2} 
\end{equation}
where 
\begin{equation}
V(t)= \int_{0}^{\infty} dt' P(t,t') e^{-iLt'}
\label{viditi} 
\end{equation}
is the evolution operator for the averaged state of the system.
Following Ref.~\cite{rb1,rb2}, we determine the function $P(t,t')$ by
imposing the following plausible
conditions: i) $\bar{\rho}(t)$ must
be a density operator, i.e. it must be self-adjoint, positive-definite, and
with unit-trace. This leads to the condition that $P(t,t')$ must be
non-negative and normalized, i.e a probability density in $t'$ so that
Eq.~(\ref{robar}) is a completely positive mapping. ii) $V(t)$ satisfies
the semigroup property $V(t_{1}+t_2)=V(t_1)V(t_2)$, with $t_1,t_2 \geq 0$.

The semigroup condition is satisfied by an exponential dependence on 
$t$
\begin{equation}
V(t) = \left\{V_{1}\right\}^{-t/\tau_{2}}
\label{semig} \;,
\end{equation} 
where $\tau_{2}$ naturally appears as a scaling time. A solution
satisfying all the conditions we have imposed can be found by
separating $V_{1}$ in its hermitian and antihermitian part $V_{1}=
A+iB$ and by considering the Gamma function integral
identity \cite{grad}
\begin{equation}
(V_1)^{-t/\tau_2}=
\left(A+iB\right)^{-t/\tau_{2}}=\frac{1}{\Gamma\left(
\frac{t}{\tau_2}\right)}\int_0 ^{\infty} d\lambda 
\lambda ^{\frac{t}{\tau_2}-1}e^{-\lambda A} e^{-i\lambda B} \;.
\label{gammaid}
\end{equation}
Now the right hand side of Eq.~(\ref{gammaid})
can be identified with the right hand side of Eq.~(\ref{viditi}) if 
we impose the following conditions:
$\lambda = t'/\tau_1 $, where $\tau_1$ is another scaling time, 
generally different from $\tau_2$; $B=L\tau_1$ in order to
make the exponential terms identical, and $A=1$ in order to get a
normalized probability distribution $P(t,t')$. 
This choice yields the following expressions for the evolution operator 
for the averaged density matrix $V(t)$ and for the probability density
$P(t,t',\tau_1,\tau_2)$
\cite{rb1,rb2}
\begin{eqnarray}
V(t) &=&\left(1+iL\tau_1\right)^{-t/\tau_2}
\label{fund} \\
P(t,t',\tau_1,\tau_2)&=& \frac{e^{-t'/\tau_1}}{\tau_1} \frac{(t'/\tau_1)^{(t/\tau_2)-1}}
{\Gamma(t/\tau_2)} \;.
\label{gamma}
\end{eqnarray}
Notice that 
the ordinary quantum evolution is recovered
when $\tau_{1}=\tau_{2}=\tau \rightarrow 0$; in this limit 
$P(t,t',\tau_{1},\tau_{2}) \rightarrow \delta(t-t')$
so that $\bar{\rho}(t)=\rho(t)$ and $V(t)=\exp\{-iLt\}$ is the 
usual unitary evolution. Moreover, it can be seen that
Eq.~(\ref{fund}) implies that 
$\bar{\rho}(t)$ satisfies a finite difference equation \cite{rb1}.
The semigroup condition leads to the form of the probability
distribution $P(t,t',\tau_1,\tau_2)$ we use to perform the average on the 
fluctuating evolution times. However, notice that this probability 
distribution depends on both the two scaling times $\tau_1$ and $\tau_2$
only apparently. In fact, if we change variable in the time integral,
$t'' =(\tau_2/\tau_1) t'$, it is possible to rewrite the integral
expression for $V(t)$ in the following way
\begin{equation}
V(t) =\left(1+iL\tau_1\right)^{-t/\tau_2}=
\int_0^{\infty} dt''
P(t,t'',\tau_2) e^{-iL(\tau_1/\tau_2)t''} \;,
\label{chvar}
\end{equation}
where
\begin{equation}
P(t,t'',\tau_2) = \frac{e^{-t''/\tau_2}}{\tau_2} \frac{(t''/\tau_2)^{(t/\tau_2)-1}}
{\Gamma(t/\tau_2)} \;.
\label{pdue}
\end{equation}
This probability density depends only on $\tau_2$. However
Eq.~(\ref{chvar}) contains an effective
{\em rescaled} time evolution generator $L_{eff} = L (\tau_1/\tau_2)$.
The physical meaning of the probability distribution of Eq.~(\ref{pdue}), 
of the
rescaled evolution operator, and of the two scaling times can be understood
if we consider the following simple example.
Let us consider a system with Hamiltonian $H(t) =f(t)H_0$,
where
\begin{equation}
f(t) = \sum_{n=0}^{\infty} \theta (t-n\tau_2) \theta(n\tau_2+\tau_1 -t)
\label{scale} 
\end{equation}
($\theta(t)$ is the Heaviside step function),
that is, a system with Hamiltonian $H_0$ which is 
periodically applied for a time $\tau_1$, with time period $\tau_2$
($\tau_2 \geq \tau_1 $)
and which is ``turned off'' otherwise. The unitary evolution operator
for this system is $U(t) = e^{-i F(t) L_0}$, where $L_0=[H_0,\ldots]$
and 
\begin{equation}
F(t) = \left\{\matrix{
 t+ n(\tau_1-\tau_2) & n\tau_2 \leq t \leq n\tau_2+\tau_1 \cr
   (n+1)\tau_1 & n\tau_2 +\tau_1 \leq t \leq (n+1)\tau_2 \cr}\right . \; , 
\label{effediti}
\end{equation}
which can be however well approximated by the ``rescaled'' evolution operator
$U_{eff}(t) = e^{-i L_0 t (\tau_1/\tau_2)}$. In fact,
the maximum relative error in replacing $F(t)$ with $t (\tau_1/\tau_2)$
is $(\tau_2 -\tau_1)/t$ and becomes negligible at large times
(see Fig.~1).
This fact suggests to interpret the time average of Eq.~(\ref{chvar}) as
an average over unitary evolutions generated by $L$, taking place
randomly in time, with mean time
width $\tau_1$, and separated by a mean time interval $\tau_2$.
This interpretation is confirmed by the fact that 
when $t = k\tau_2$, for integer $k$, the probability distribution
$P(t,t'',\tau_2)$ of Eq.~(\ref{pdue}) is a known statistical
distribution giving the probability density that the waiting time 
for $k$ independent events
is $t''$ when $\tau_2$ is the mean time interval between two events.
A particularly clear example of the random process in time
implied by the above equations is provided by the micromaser 
\cite{micro} in which a microwave cavity is crossed by a beam 
of resonant atoms with mean injection rate $R=1/\tau_{2}$, and a mean 
interaction time within the cavity corresponding to $\tau_{1}$. In 
the micromaser theory, the non unitary operator $M$ describing the 
effective dynamics of the microwave mode during each atomic crossing
replaces the evolution operator $e^{-iL\tau_{1}}$ \cite{rb1}.
Another example of interrupted evolution is provided by the 
experimental scheme proposed in \cite{brune} for the quantum 
non-demolition (QND) measurement \cite{caves} of the photon number
in a high-Q cavity. In this proposal, the photon number is determined
by measuring the phase shift induced on a train of Rydberg atoms 
sent through the microwave cavity with
mean rate $1/\tau_{2}$, and interacting dispersively with the cavity mode.
These two examples show that
the two scaling times $\tau_1$ and $\tau_2$ have not to be 
considered as new universal constants, but as two characteristic times 
of the system under study. 

However, in most cases, one does not have an interrupted evolution as 
in micromaser-like situations, but a standard, 
continuous evolution generated
by an Hamiltonian $H$. In this case the ``scaled'' effective evolution
operator has to coincide with the usual one, $L$, and this is 
possible only if $\tau_{1}=\tau_{2} = \tau$. In this case 
$\tau$ is simply the parameter characterizing the strength of the
fluctuations of the random evolution time. This meaning of the
parameter $\tau $ in the case of equal scaling times is confirmed
by the expressions of the mean and the variance of the
probability distribution of Eq.~(\ref{gamma})
\begin{eqnarray}
\langle t' \rangle & = & \frac{\tau_{1}}{\tau_{2}}t \\
\sigma^{2}(t') &=& \langle t'^{2}\rangle -\langle t'\rangle ^{2} =
 \frac{\tau_{1}^{2}}{\tau_{2}}t\;.
\end{eqnarray}
When $\tau_{1}=\tau_{2}$, the mean evolution time coincide with
the ``clock `' time $t$, while the variance of the evolution time
becomes $\sigma^{2}(t')=t\tau $. In the rest of the paper
we shall always consider the standard situation of an isolated
system with Hamiltonian $H$, continuously evolving in time, and we
shall always assume $\tau_{1}=\tau_{2}=\tau$.

When $\tau = 0$, $V(t)=\exp\{-iLt\}$ is the 
usual unitary evolution. For finite $\tau$, on the contrary,
the evolution equation
(\ref{fund}) describes a decay of the off-diagonal matrix elements
in the energy representation, whereas the diagonal matrix elements remain
constant, i.e. the energy is still a constant of motion.
In fact, in the energy eigenbasis, Eqs.~(\ref{robar2}) and (\ref{fund})
yield
\begin{equation}
	\bar{\rho}_{n,m}(t)= \frac{1}{(1+i\omega_{n,m}\tau_{1})^{t/\tau_{2}}}
	\rho_{n,m}(0)=
	\frac{e^{-i\nu_{n,m}t}}{(1+\omega_{n,m}^{2}
	\tau_{1}^2)^{t/2\tau_{2}}}=e^{-\gamma_{n,m}t}e^{-i\nu_{n,m}t}
	\rho_{n,m}(0) \;,
	\label{elematri}
\end{equation}
where $\omega_{n,m}= (E_n-E_m)/\hbar$ and
\begin{eqnarray}
\gamma_{n,m}&=&\frac{1}{2\tau_2} \log\left(1+\omega_{n,m}^2\tau_1^2\right)
\label{gammamn} \\
\nu_{n,m}&=&\frac{1}{\tau_2}\arctan \left(\omega_{n,m}\tau_1\right)
\label{numn} \;.
\end{eqnarray}
This means that, in general, the effect of the average over the 
fluctuating evolution time yields an exponential decay 
and a frequency shift $\omega_{n,m} \rightarrow \nu_{n,m}$
of every term oscillating in time with frequency $\omega_{n,m}$.

The phase diffusion aspects of the present approach can also be seen
if the evolution equation of the averaged density matrix $\bar{\rho}(t)$
is considered.
In fact, by differentiating with respect to time Eq.~(\ref{robar2})
and using (\ref{fund}),
one gets the following master equation for $\bar{\rho}(t)$
(we consider the case $\tau_1=\tau_2=\tau$)
\begin{equation}
\dot{\bar{\rho}}(t)=-\frac{1}{\tau}\log\left(1+iL\tau\right) \bar{\rho}(t)
\label{ropun} \;;
\end{equation}
expanding the logarithm at second order in $L\tau$, one obtains
\begin{equation}
\dot{\bar{\rho}}(t)=
-\frac{i}{\hbar}\left[H,\bar{\rho}(t)\right]-\frac{\tau}{2\hbar^2}\left[
H,\left[H,\bar{\rho}(t)\right] \right] \;,
\label{ropun2}
\end{equation}
which is the well-known phase-destroying master equation \cite{milwal}.
Hence Eq.~(\ref{ropun}) appears as a {\em generalized} phase-destroying
master equation taking into account higher order terms in $\tau$.  
Notice, however, that the present approach is different from the
usual master equation approach in the sense that it is 
model-independent and no perturbative and specific
statistical assumptions are made.
The solution of Eq.~(\ref{ropun2}) gives an expression for
$\bar{\rho}_{n,m}(t)$ similar to that of Eq.~(\ref{elematri}), but
with \cite{milwal}
\begin{eqnarray}
\gamma_{n,m}&=&\frac{\omega_{n,m}^2\tau}{2} 
\label{gammamn2} \\
\nu_{n,m}&=&\omega_{n,m}
\label{numn2} \;,
\end{eqnarray}
which are nonetheless obtained also as a first order expansion
in $\tau_1=\tau_2=\tau$ of Eqs.~(\ref{gammamn}) and (\ref{numn}).
The opposite limit $\omega_{m,n}\tau \gg 1$ has been discussed in detail
in Ref.~\cite{rb1}.

Finally a comment concerning the form of the evolution operator
for the averaged density matrix $V(t)$ of Eq.~(\ref{fund}).
At first sight it seems that $V(t)$ is in general a multivalued function of
the Liouvillian $L$, and that $V(t)$ is uniquely defined only when
$t/\tau_{2}=k$,
$k$ integer. However, this form for $V(t)$ is a consequence of the 
time average over 
$P(t,t',\tau_{1},\tau_{2})$ of Eq.~(\ref{gamma}),
which is a properly defined, non-negative
probability distribution only if the algebraic definition of the 
power law function $(t'/\tau_{1})^{(t/\tau_{2})-1}$ is assumed.
This means that in Eq.~(\ref{fund}) one has to take the
first determination of the power-law function and in this way
$V(t)$ is univocally defined.

\section{Application to cavity QED experiments} 

A first experimental situation in which the above formalism 
can be applied is the Rabi oscillation experiment
of Ref.~\cite{haro}, in which the 
resonant interaction between a quantized mode in a high-Q microwave 
cavity (with annihilation operator $a$)
and two circular Rydberg states ($|e \rangle $ and $|g \rangle $)
of a Rb atom has been studied. This 
interaction is well described by the usual Jaynes-Cummings \cite{jc}
model, which in the interaction picture reads
\begin{equation}
H=\hbar \Omega_{R} \left(|e \rangle \langle g |a+|g \rangle \langle e |
a^{\dagger}\right) \;,
\label{jc}
\end{equation}
where $\Omega_{R}$ is the Rabi frequency.

The Rabi oscillations describing the exchange of excitations
between atom and cavity mode are studied by injecting
the velocity-selected Rydberg atom, prepared
in the excited state $|e \rangle $, in the high-Q cavity and measuring the
population of the lower atomic level $g$,
$P_{eg}(t)$, as a function of the interaction time $t$, which is varied 
by changing the Rydberg atom velocity. 
Different initial states of the cavity mode have been considered in
\cite{haro}. We shall restrict only to the case of vacuum state 
induced Rabi oscillations, where the decoherence effect is particularly
evident. The
Hamiltonian evolution according to Eq.~(\ref{jc}) predicts in this case
Rabi oscillations of the form 
\begin{equation}
P_{eg}(t) =\frac{1}{2} \left(1- \cos \left(2\Omega_{R} 
t\right)\right)\;.
\label{rabi1}
\end{equation}
Experimentally instead, damped oscillations are observed, which are well
fitted by 
\begin{equation}
P_{eg}^{exp}(t) =\frac{1}{2} \left(1- e^{-\gamma t}
\cos \left(2\Omega_R t\right)\right) \;,
\label{fit}
\end{equation}
where the decay time fitting the experimental data
is $\gamma ^{-1} =40 \mu$sec \cite{pri} and the corresponding 
Rabi frequency is $\Omega_R/2\pi =25$ Khz (see Fig.~2). This decay of quantum 
coherence cannot be associated with photon leakage out of the cavity
because the cavity relaxation time is larger ($220$ $\mu$sec) 
and also because in this case one would have an asymptotic limit
$P_{eg}^{exp}(\infty)=1$. Therefore decoherence in this case
has certainly a non dissipative origin, and dark counts of 
the atomic detectors, dephasing collisions with background gas or
stray magnetic fields within the cavity
have been suggested as possible sources of the damped
oscillations. \cite{haro,pri}.

The damped behavior of Eq.~(\ref{fit})
can be easily obtained if one applies the formalism described above.
In fact, from
the linearity of Eq.~(\ref{robar}), 
one has that the
time averaging procedure is also valid for mean values and matrix
elements of each subsystem. Therefore one has
\begin{equation}
\bar{P}_{eg}(t) = \int_{0}^{\infty} dt' P(t,t') P_{eg}(t') \;.
\label{pegbar} 
\end{equation}
Using Eqs.~(\ref{robar2}), (\ref{fund}), (\ref{gamma}) and (\ref{rabi1}),  
Eq.~(\ref{pegbar}) can be rewritten in the same form 
of Eq.~(\ref{fit})
\begin{equation}
\bar{P}_{eg}(t) =\frac{1}{2} \left(1- e^{-\gamma t}
\cos \left(\nu t\right)\right) \;,
\label{fit2}
\end{equation}
where, using Eqs.~(\ref{gammamn}) and (\ref{numn}), 
\begin{eqnarray}
\label{gammalog}
\gamma &= &\frac{1}{2\tau} \log \left(1+4\Omega_{R}^2 \tau^2\right) \\
\nu &= &\frac{1}{\tau} {\rm arctg}\left(2\Omega_R \tau\right) \;.
\end{eqnarray}
If the characteristic time $\tau$
is sufficiently small, i.e. $\Omega_R \tau \ll 1$, there is
no phase shift, $\nu \simeq 2\Omega_R$,
and 
\begin{equation}
\gamma = 2 \Omega_{R}^{2}\tau 
\label{gammapr}
\end{equation}
(see also Eqs.~(\ref{gammamn2}) and (\ref{numn2})).
The fact that in Ref.~\cite{haro} the Rabi oscillation frequency essentially
coincides with the theoretically expected one, suggests that the time 
$\tau$ characterizing the fluctuations of the interaction time is 
sufficiently small so that it is reasonable to use Eq.~(\ref{gammapr}).
Using the above values for $\gamma$ and $\Omega_{R}$,
one can derive an estimate for $\tau$, so to
get $\tau \simeq 0.5$ $\mu$sec. This estimate is consistent with 
the assumption $\Omega_{R}\tau \ll 1$ we have made, but,
more importantly, it turns out to be
comparable to the experimental value of the uncertainty in the interaction time.
In fact, the fluctuations of the interaction time are mainly due to the 
experimental uncertainty
of the atomic velocity $v$, that is 
$ \delta t/t \simeq \delta v/v =1\%$ (see Ref.~\cite{haro}), and taking
an average interaction time $\bar{t} \simeq 50$ $\mu$sec, one gets
 $\tau \simeq \delta t
= \bar{t} \delta v/v =0.5$ $\mu$sec, 
which is just the estimate we have derived from the 
experimental values.
This simple argument supports the interpretation that
the decoherence observed in \cite{haro} is essentially
due to the randomness of the interaction time. In fact, in our opinion,
the other effects proposed as possible sources of decoherence, such
as dark counts of 
the atomic detectors, dephasing collisions with background gas or
stray magnetic fields within the cavity, would give an overall, 
time-independent, contrast reduction of the Rabi oscillations, 
different from the observed exponential decay.

Results similar to that of Ref.~\cite{haro} have been very recently
obtained by H. Walther group at the Max Planck Institut f\"ur Quantenoptik,
in a Rabi oscillation experiment involving again a high-Q microwave
cavity mode resonantly interacting with Rydberg atoms \cite{pri3}.
In this case, three different initial Fock states $|n\rangle $
of the cavity mode, 
$n=0,1,2$, have been studied, and preliminary results show a good
quantitative agreement of the experimental data with our
theoretical approach based on the dispersion of the interaction times.

Another cavity QED experiment in which the observed
decay of quantum coherence
can be, at least partially, explained with our
formalism in terms of a random interaction time,
is the Ramsey interferometry
experiment of M. Brune {\it et al.}
\cite{disp}. In this experiment, a 
QND measurement of the mean photon
number of a microwave cavity mode is obtained
by measuring, in a Ramsey interferometry scheme, the dispersive 
light shifts produced on circular
Rydberg states by a nonresonant microwave field.
The experimental scheme in this case is similar to that of 
the Rabi oscillation experiment, with two main differences:
i) two low-Q microwave cavities $R_1$ and $R_2$, which can be fed by
a classical source $S$ with frequency $\omega_R$, are added just before
and after the cavity of interest $C$; 
ii) the cavity mode is highly detuned 
from 
the atomic transition ($\delta =\omega-\omega_{eg} \gg \Omega_R $), 
so to work in the dispersive regime.
In the interaction picture with respect to
$$H_0 = \frac{\hbar \omega_{R}}{2}\left[|e\rangle \langle e | -
|g\rangle \langle g |\right] $$
(we use
the classical field as reference for the atomic phases),
the Hamiltonian has the following dispersive form \cite{brune}
\begin{eqnarray}
H&=&\frac{\hbar \Delta}{2}\left[|e\rangle \langle e | -
|g\rangle \langle g |\right] +\hbar \chi (t) \left(|g \rangle \langle e |+
|e\rangle \langle g | \right) \\
&&+ \hbar \omega a^{\dagger} a +
\hbar \frac{\Omega_{R}^2(t)}{\delta} \left(|g \rangle \langle g |a^{\dagger}
a - |e\rangle \langle e |
aa^{\dagger}\right)  \;,
\label{disper}
\end{eqnarray}
where $\Delta = \omega_{eg}-\omega_R$, the Rabi frequency
within the classical cavities
$\chi(t) $ is nonzero only when the atom is in $R_{1}$ and $R_{2}$,
and $\Omega_R(t)$ is nonzero only within $C$.
In the experiment, single circular Rydberg atoms are sent through the  
apparatus initially prepared in the state $|e\rangle $, and let us assume
that the microwave cavity mode in $C$ is in a generic state
$\sum_n c_n |n\rangle $.
The atom is subject to a $\pi/2$ pulse in $R_1$, so that
\begin{equation}
\sum_n c_n |n\rangle |e\rangle \rightarrow \sum_n c_n |n\rangle 
\frac{\left(|e\rangle + |g\rangle \right)}{\sqrt{2}}
\label{firstpi} \;.
\end{equation}
Then the atom crosses the cavity $C$ with an interaction
time $t_{int}$
and the dispersive interaction yields
\begin{equation}
|\psi\rangle = \sum_n \frac{c_n e^{-i \omega n t_{int}}}{\sqrt{2}}|n\rangle 
\left(e^{-i \frac{\Delta}{2} t_{int}+i
\frac{\Omega_R^2}{\delta}(n+1) t_{int}}|e\rangle + 
e^{i \frac{\Delta}{2} t_{int}-i
\frac{\Omega_R^2}{\delta}n t_{int}}|g\rangle \right) \;.
\label{dispc}
\end{equation}
Finally the atom is subject to the second
$\pi/2$ pulse in the second Ramsey zone $R_2$ and the joint 
state of the Rydberg atom and the cavity mode becomes
\begin{eqnarray}
&&|\psi\rangle = \sum_n c_n e^{-i \omega n T+
i\frac{\Omega_R^2}{2\delta}t_{int}}|n\rangle 
\left\{\cos\left[ \frac{\Omega_R^2}{\delta}(n+\frac{1}{2}) t_{int}
-\frac{\Delta}{2} T\right]|g\rangle \right.  \\
&& \left.+ i
\sin\left[ \frac{\Omega_R^2}{\delta}(n+\frac{1}{2}) t_{int}
-\frac{\Delta}{2} T\right]|e\rangle \right\} \;, 
\label{statofin}
\end{eqnarray}
where $T$ is the time of flight from $R_1$ to $R_2$.
The experimentally interesting quantity is
the probability to find at the end the atom in the $g$ state, 
$P_{eg}(n,T)$, whose theoretical expression according to
Eq.~(\ref{statofin}) is
\begin{equation}
P_{eg}(n,T)=\cos^2\left[\frac{T}{2}\left(\Delta -\epsilon_n\right)
\right] \;,
\label{peg}
\end{equation}
where the photon number-dependent frequency shift $\epsilon_n$ is
given by
\begin{equation}
\epsilon_n =\frac{\Omega_R^2}{\delta}
\frac{w}{d(R_1,R_2)}(2n+1) \;.
\label{freshift}
\end{equation}
In Eqs.~(\ref{peg}) and (\ref{freshift})
we have used the fact that
$t_{int}/T$ is equal to the ratio between the
waist of the cavity mode $w$ and the distance between the two
Ramsey cavities $d(R_1,R_2)$. 
The actual experiment of Ref.~\cite{disp} has been
performed in the bad cavity limity $T_{rel} < t_{int} $ in which the
cavity $C$ relaxation time $T_{rel}$ is smaller than the atom-cavity
interaction time. In this case, the cavity photon number
randomly changes during $t_{int}$ and in the corresponding expression
(\ref{freshift}) for the frequency shift $\epsilon_{n}$, 
the photon number $n$ has to replaced by the mean value $\bar{n}$.
The Ramsey fringes are observed by sweeping the frequency of the classical
source $\omega_R$ around resonance, that is, studying $P_{eg}(\bar{n},T)$
as a function of the detuning $\Delta$.
The experimentally observed Ramsey fringes show a reduced contrast, which
moreover decreases for increasing detunings $\Delta$ (see Fig.~2 of 
Ref.~\cite{disp}). Therefore one can try to explain the reduced
contrast, i.e., the loss of quantum coherence, in terms of a fluctuating
evolution time, which in this case means a random time of flight $T$
originated again by the dispersion of the atomic velocities.
We average again the quantity $P_{eg}$ of Eq.~(\ref{peg})
over the probability
distribution $P(t,t')$ derived in Section II, replacing $t'$ with
a random time of flight $T'$, and we obtain
\begin{equation}
\bar{P}_{eg}(\Delta)=\frac{1}{2}
\left\{1+F(\Delta-\epsilon_n)\cos\left[\left(\Delta ' -\epsilon_n '\right)T
\right]\right\} 
\label{pegmedio}
\end{equation}
where, using Eq.~(\ref{elematri}), 
the fringe visibility function $F(\Delta -\epsilon_n) $
is given by
\begin{equation}
F(\Delta -\epsilon_n)=\left[1+(\Delta-\epsilon_n)^2 \tau ^2\right]^{-T/2\tau}
\label{fringe}\;,
\end{equation}
and $\left(\Delta ' -\epsilon_n '\right) $ is the frequency shift
\begin{equation}
\left(\Delta ' -\epsilon_n '\right) = \frac{1}{\tau}{\rm arctg}\left[
\left(\Delta -\epsilon_n\right)\tau\right] 
\label{freshift2}\;.
\end{equation}
The parameter $\tau$ characterizing the strength of the fluctuations
of the time of flight $T$ can be estimated with arguments similar to
those considered for the Rabi oscillation experiment.
Since $\delta T /T \simeq \delta v/v =1.5 \%$ and $T\simeq 300$
$\mu$sec (see Ref.~\cite{disp}), one has $\tau \simeq \delta T \simeq
4.5$ $\mu$sec. For the interesting range of detunings $\Delta $,
one has $ \left(\Delta -\epsilon_n\right)\tau \ll 1 $, so that
one can neglect again the frequency shift (\ref{freshift2})
and approximate the fringe visibility function (\ref{fringe})
with a gaussian function, that is,
\begin{equation}
\bar{P}_{eg}(\Delta)=\frac{1}{2}
\left\{1+e^{-\frac{\left(\Delta-\epsilon_n\right)^2}{2} T\tau}
\cos\left[\left(\Delta -\epsilon_n \right)T
\right]\right\} \;.
\label{gaumodu}
\end{equation}
This gaussian modulation of the Ramsey fringes
with a width $\sigma_{\Delta} = (T\tau)^{-1/2} \simeq 27 \; {\rm Khz}$
is consistent with the typical experimental Ramsey fringe signal
(see Fig.~2 of Ref.~\cite{disp}), but it is not able to
completely account for the observed modulation and contrast reduction
of the fringes. This means that, contrary to the
case of the Rabi oscillation experiment, in this case
the role of other experimental imperfections such as
random phases due to stray fields, imperfect $\pi/2 $ pulses
in $R_1$ and $R_2$ and detection errors, is as relevant as 
that of the dispersion of atomic velocities and these other effects 
have to be taken into account to get
an exhaustive explanation of the observed decoherence.

\section{Rabi oscillation experiments in trapped ions} 

Another interesting Rabi oscillation experiment,
performed on a different system, that is, a trapped ion
\cite{wine}, has recently observed a decoherence effect
which cannot be attributed to dissipation.
In the trapped ion experiment of Ref.~\cite{wine}, the 
interaction between two internal states ($|\uparrow \rangle
$ and $|\downarrow \rangle$) of a Be ion and the 
center-of-mass vibrations in the $z$ direction, induced by two
driving Raman lasers is studied. 
In the interaction picture with respect to the free vibrational and 
internal Hamiltonian, this interaction is described by the following
Hamiltonian \cite{nist}
\begin{equation}
H = \hbar \Omega |\uparrow \rangle \langle \downarrow |
\exp\left\{i\left[\eta \left(
a e^{-i\omega_{z}t}+a^{\dagger}e^{i\omega_{z}t}\right)-\delta t+\phi
\right]\right\} +  H.C. \;,
\label{hgen}
\end{equation}
where $a$ denotes the 
annihiliation operator for the vibrations along the $z$ direction, 
$\omega_{z}$ is the corresponding frequency and $\delta $ 
is the detuning between the 
internal transition and the frequency difference between the two Raman 
lasers. The Rabi frequency $\Omega$ is proportional to the 
two Raman laser intensities, and $\eta $ is the Lamb-Dicke parameter
\cite{wine,nist}.
When the two Raman lasers
are tuned to the first blue sideband, i.e. $\delta = 
\omega_{z}$, 
Hamiltonian (\ref{hgen}) predicts Rabi oscillations between $|\downarrow,n\rangle $
and $|\uparrow, n+1\rangle$ ($|n\rangle $ is a vibrational
Fock state) with a frequency \cite{nist}
\begin{equation}
\Omega_n = \Omega \frac{e^{-\eta^2/2}}{\sqrt{n+1}}\eta L_{n}^{1}(\eta ^2) \;,
\label{omn}
\end{equation}
where $L_{n}^{1}$ is the generalized Laguerre polynomial.
These Rabi oscillations have been experimentally verified
by preparing the initial state $|\downarrow , n
\rangle $, (with 
$n$ ranging from $0 $ to $16$) and measuring the probability
$P_{\downarrow}(t)$ as a function of the interaction time $t$, which is varied
by changing the duration of the Raman laser pulses. 
Again, as in the cavity QED experiment of \cite{haro}, 
the experimental Rabi oscillations are damped and well fitted by \cite{wine,nist}
\begin{equation}
P_{\downarrow}(n,t) =\frac{1}{2} \left(1+ e^{-\gamma_n t}
\cos \left(2\Omega_n t\right)\right) \;,
\label{fitwine}
\end{equation}
where the 
measured oscillation frequencies $\Omega_n$ are in very good agreement with 
the theoretical prediction (\ref{omn})
corresponding to the measured Lamb-Dicke parameter $\eta =0.202$ \cite{wine}.
As concerns the decay rates $\gamma_n$, 
the experimental values are fitted in \cite{wine} by
\begin{equation}
\gamma_n = \gamma_0 (n+1)^{0.7}
\label{o7}
\end{equation}  
where $\gamma_0=11.9$ Khz. This power-law scaling 
has attracted the interest of a number of authors
and it has been investigated in Refs.~\cite{milb,murao}, even if
a clear explanation of this behavior of the decay rates is still lacking.
On the contrary, the scaling law (\ref{o7}) can be simply accounted for
in the previous formalism if we consider the small $\tau$ limit
of Eq.~(\ref{gammapr}), which is again suggested by the
fact that the experimental and theoretical predictions for the
frequencies $\Omega_n$ agree. 
In fact, the $n$-dependence of the theoretical prediction
of Eq.~(\ref{omn}) for $\eta =0.202$ is well approximated, within 10 \%,
by the power law dependence (see Fig.~3) 
\begin{equation}
\Omega_n \simeq \Omega_0 (n+1)^{0.35} \;,
\label{o35}
\end{equation}  
so that, using Eq.~(\ref{gammapr}), one has immediately the power law dependence
$(n+1)^{0.7}$ of Eq.~(\ref{o7}).
The value of the parameter $\tau$ can be obtained by matching the
values corresponding to $n=0$, and using Eq.~(\ref{gammapr}), that is
$\tau = \gamma_0/2\Omega_0 ^2 \simeq 1.5 \cdot 10^{-8}$ sec,
where we have used the experimental value $\Omega_{0}/2\pi = 94$ Khz.

However, this value of the parameter $\tau$ cannot be explained in terms
of some interaction time uncertainty, such as the time jitter 
of the Raman laser pulses, which is experimentally found to be
much smaller \cite{pri2}. In this case, instead, the observed decoherence
can be attributed,  as already suggested in \cite{nist,milb,murao}, 
to the fluctuation of the Raman laser
intensities, yielding a fluctuating Rabi frequency parameter
$\Omega (t)$ of the Hamiltonian (\ref{hgen}). In this case,
the evolution is driven by a fluctuating Hamiltonian $H(t)=
\hbar \Omega(t) \tilde{H}$, where $\tilde{H}=H/\Omega$ in Eq.~(\ref{hgen}), 
so that
\begin{equation}
\rho(t) = \exp\{-i \tilde{L}\int_{0}^{t}d\xi \Omega(\xi)\}\rho(0)
=e^{-i\tilde{L} A(t)} \rho(0)
\label{rotild}
\end{equation}
where $\tilde{L}=[\tilde{H},\ldots ]/\hbar$, and we have defined
the positive dimensionless random variable $A(t)=\int_0 ^t d\xi \Omega(\xi)$, 
which is proportional to the pulse area. It is now easy to understand
that the physical situation is analogous to that characterized by
a random interaction time considered in the preceding
sections, with $L$ replaced
by $\tilde{L}$ and $t'$ by $A(t)$. It is therefore
straightforward to adapt the formalism developed in Section II
to this case, in which the fluctuating quantity is the pulse area $A$,
yielding again random phases in the energy basis representation.
In analogy with
Eq.~(\ref{robar}), one considers 
an averaged density matrix
\begin{equation}
\bar{\rho}(t)= \int_{0}^{\infty} dA P(t,A) e^{-i\tilde{L} A} \rho(0) \;.
\label{robara} 
\end{equation}
Imposing again that $\bar{\rho}(t) $
must be a density operator and the semigroup property, one finds 
results analogous to Eqs.~(\ref{fund}) and (\ref{gamma})
\begin{eqnarray}
V(t) &=&\left(1+i\tilde{L}\Omega\tau\right)^{-t/\tau}
\label{funda} \\
P(t,A)&=& \frac{e^{-A/\Omega\tau}}{\Omega\tau} 
\frac{(A/\Omega\tau)^{(t/\tau)-1}}
{\Gamma(t/\tau)} \;.
\label{agamma}
\end{eqnarray}
Here, the parameters $\Omega$ and $\tau$ are introduced as
scaling parameters, but they have a clear meaning, as it can be
easily seen by considering the mean and the variance of
the probability distribution
of Eq.~(\ref{agamma}), 
\begin{eqnarray}
&& \langle A \rangle =\Omega t \\
&& \sigma^2(A)=\langle A^2\rangle - \langle A \rangle ^2 = \Omega^2 t\tau 
\end{eqnarray}
implying that $\Omega$ has now to be meant as a {\em mean} 
Rabi frequency, and that $\tau $ quantifies the strength of $A$ fluctuations.
It is interesting to note that
these first two moments of $P(t,A)$ determine the
properties of the fluctuating Rabi frequency $\Omega(t)$, which can be
written as
\begin{eqnarray}
&&\Omega(t) = \Omega + \xi(t) \\
&& \langle \xi(t) \rangle =0
\;\;\;\; \langle \xi(t) \xi(t') \rangle = \Omega^2 \tau \delta(t-t')
\;, 
\end{eqnarray}
that is, the Rabi frequency $\Omega(t)$ is a white, non-gaussian (due to
the non-gaussian form of
$P(t,A)$) stochastic process. In fact, the semigroup 
assumption we have made implies a Markovian treatment 
in which the spectrum of the laser intensity fluctuations
is flat in the relevant frequency range.
This in particular implies that
we are neglecting the dynamics at small times, of the order of the
correlation time of the laser intensity fluctuations.

The estimated value of $\tau$ gives a reasonable estimate
of the pulse area fluctuations,
since it corresponds to a fractional error of the pulse area
$\sqrt{\sigma^2(A)}/\langle A \rangle = \sqrt{\tau/t}$ of $10 \%$ 
for a pulse duration of $t=1$ $\mu$sec, and which is decreasing for
increasing pulse durations.

The present analysis shows many similarities with that of Ref.~\cite{milb}
which also tries to explain the decay of the Rabi oscillations in the ion trap
experiments of \cite{wine} in terms of laser intensity fluctuations.
The authors of Ref.~\cite{milb} in fact use a phase 
destroying master equation coinciding with the second-order expansion
(\ref{ropun2}) of our generalized master equation of Eq.~(\ref{ropun})
(see Eq.~(16) of Ref.~\cite{milb} with the identifications $G \leftrightarrow
H/\hbar$ and $\Gamma \leftrightarrow \tau$) and moreover derive the
same numerical estimate for the pulse area fluctuation strength 
$\Gamma \leftrightarrow \tau$. Despite this similarities, 
they do not recover the scaling (\ref{o7}) of the decay rates 
$\gamma_n$ only because they
do not use the general expression of the Rabi frequency (\ref{omn}),
(and which is well approximated by the power law (\ref{o35}))
but its Lamb-Dicke limit
$ \Omega_n = \Omega_0 (n+1)^{0.5} $,
which is valid only when $\eta \ll 1$.
There is however another, more fundamental, difference between 
our approach and that of Ref.~\cite{milb}. They
assume from the beginning that the laser intensity fluctuations have
a white and gaussian character, while we make no {\em a priori}
assumption on the statistical properties of the pulse area $A$.
We derive these properties, i.e. the probability distribution (\ref{agamma}),
only from the semigroup condition, and it is interesting to note
that this condition yields a gaussian probability distribution
for the pulse area only as a limiting case. In fact, from Eq.~(\ref{agamma})
one can see that $P(t,A)$ tends to become a gaussian with the same
mean value
$\Omega t$ and the same width $\Omega ^2 \tau t$ only 
in the large time limit
$t/\tau \gg 1$
\begin{equation}
P(t,A)_{t\gg \tau} \simeq \frac{1}{\sqrt{2\pi \Omega^2 t \tau}}
\exp\left\{-\frac{(A-\Omega t)^2}{2\Omega^2 t \tau}
\right\}
\label{gauap}
\end{equation}
The non-gaussian character of $P(t,A)$
can be traced back to the fact that 
$P(t,A)$ must be definite
and normalized in the interval $0<A<\infty$ and not in
$-\infty < A < +\infty$. Notice that at $t=\tau$, Eq.~(\ref{agamma})
assumes the exponential form $P(t,A)=e^{-A/\Omega \tau}/\Omega \tau$.
Only at 
large times $t$ the random variable $A$ becomes the sum of many independent
contributions and assumes the gaussian form. 

Due to the non-gaussian nature of the random variable $A$, we find that
the more generally valid phase-destroying master equation is given by
Eq.~(\ref{ropun}) (with $L$ replaced by $\Omega\tilde{L}$). 
The predictions of Eq.~(\ref{ropun}) significantly depart from its
second order expansion in $L\tau$, Eq.~(\ref{ropun2}), corresponding
to the gaussian limit, as soon as $\tau$
becomes comparable with the typical timescale of the system under study, 
which, in the present case, is the inverse of the Rabi frequency.

The present analysis of the Rabi oscillation experiment
of Ref.~\cite{wine} can be repeated for the very recent experiment
with trapped ions performed in Innsbruck \cite{blatt},
in which Rabi oscillations involving the vibrational
levels and an optical quadrupole transition
of a single $^{40}$Ca$^+$ ion have been observed.
Damped oscillations corresponding to initial vibrational numbers
$n=0$ and $n=1$ are reported. From the data with $n=0$, 
$\Omega_0/2\pi = 21$ Khz and $\gamma_0 = 1$ Khz, 
we get $\tau \simeq \gamma_0/2\Omega_0^2 \simeq 3\cdot 
10^{-8} \; {\rm sec}$ and
this estimate is consistent with attributing again the decoherence  
to the fluctuations of the Rabi frequency 
caused by laser intensity fluctuations.
Moreover in this case, the experiment is performed in the Lamb-Dicke
limit $\eta \ll 1$, and therefore, using again Eq.~(\ref{gammapr}), 
we expect, in this case, a linear scaling with the vibrational
number, $\gamma_n = 2 \Omega_n^2 \tau \simeq \gamma_0 (n+1) $.

\section{Concluding remarks}

Decoherence is not always necessarily due to the entanglement with
an environment, but it may be due, as well, to the fluctuations of 
some classical parameter or internal variable of a system. 
This is a different form of decoherence, which 
is present even in isolated systems, and that we have
called non-dissipative decoherence.
In this paper we have presented a model-independent theory
for non-dissipative decoherence, which can be applied in the 
case of a random evolution time or in the case of a fluctuating
Hamiltonian. This approach 
proves to be a flexible tool, able to give
a quantitative understanding of the decoherence caused
by the fluctuations of classical quantities.
In fact, in this paper we have given a simple and 
{\em unified} description of the decoherence phenomenon
observed in recent Rabi oscillation experiments performed
in a cavity QED configuration \cite{haro} and on a trapped ion 
\cite{wine}. In particular, this approach has allowed us
to explain for the first time in simple terms, the power-law scaling of the
coherence decay rates of Eq.~(\ref{o7}), observed in the 
trapped ion experiment.

The relevant aspect of the approach applied here,
and introduced in Ref.~\cite{rb1}, is its model-independence.
The formalism is in fact derived starting from few, very general assumptions:
i) the average density matrix $\bar{\rho}(t)$
has all the usual properties of a density matrix; 
ii) the semigroup property for the time evolution generator
$V(t)$ for $\bar{\rho}(t)$.
With this respect, this approach seems to provide a very general
description of non-dissipative decoherence, in which
the random properties of the fluctuating classical variables are
characterized by the two, system-dependent, time parameters
$\tau_1$ and $\tau_2$. As we have seen in section II, in the cases 
where one has a standard, continuous evolution, the two times coincide
$\tau_1=\tau_2=\tau$. Under ideal conditions of no fluctuating 
classical variable or parameter, one would have $\tau=0$, and the usual unitary
evolution of an isolated system in quantum mechanics would be
recovered. However, the generality of the approach suggests in some
way the possibility that the parameter $\tau$, even though system-dependent,
might have a lower {\em nonzero} limit, which would be reached
just in the case of no fluctuations of experimental origin.
This would mean a completely new description of time
in quantum mechanics. In fact, the evolution time of a system $t'$
(and not 
the ``clock'' time $t$) would become an intrinsically random variable 
with a well 
defined probability distribution, {\em without} 
the difficulty of introducing an 
evolution time operator. In Ref.~\cite{rb1} it is suggested a 
relation of the nonzero limit for $\tau $ with the ``energy-time'' $\hbar 
/2\Delta E,$ where $\Delta E$ is the uncertainty in energy. This would 
give a precise meaning to the time-energy uncertainty relation because 
now $\tau $ rules the width of the time distribution function. 
However, this ``intrinsic 
assumption'' is not
necessarily implied by the formalism
developed in \cite{rb1} and applied, with a more pragmatic attitude,
in the present paper.

\section{Acknowledgments}

This work has been partially supported by INFM through 
the PAIS ``Entanglement and decoherence''.
Discussions with J.M. Raimond, H. Walther, and D. Wineland are greatly
acknowledged.

\bibliographystyle{unsrt}

\begin{figure}
\centerline{\epsfig{figure=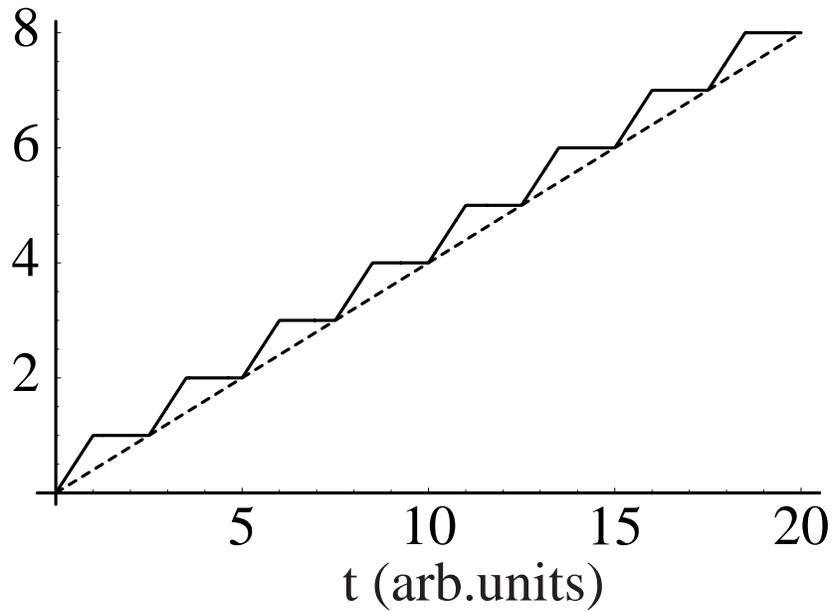,width=12cm}}
\caption{The function $F(t)$ defined in Eq.~(\protect\ref{effediti}) 
(full line) is
plotted as a function of time (expressed in arbitrary units) and
compared with its ``linear approximation'', the rescaled time
$t\tau_1/\tau_2$ (dashed line). The relative error between them
is given by $(\tau_2-\tau_1)/t$ and is negligible at large times $t$.}
\label{scalette}
\end{figure}

\begin{figure}
\centerline{\epsfig{figure=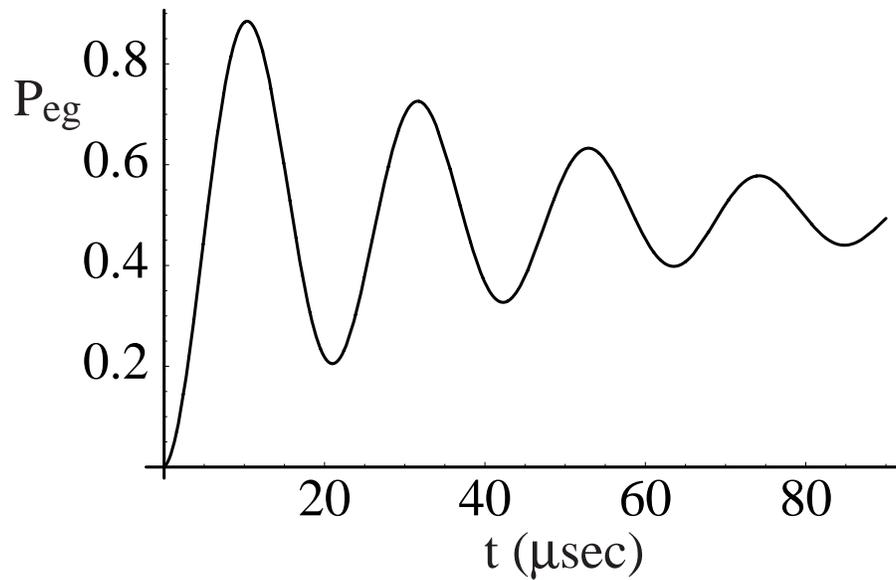,width=12cm}}
\caption{The Rabi oscillations of the transition probability
$P_{eg}(t)$ as a function of time,
according to the fitting function of Eq.~(\protect\ref{fit}).}
\label{rabiens}
\end{figure}

\begin{figure}
\centerline{\epsfig{figure=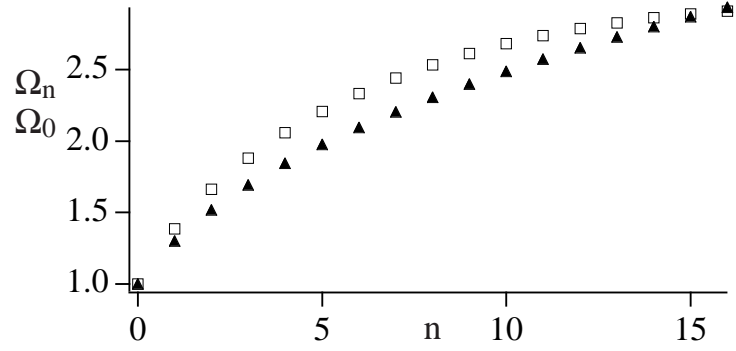,width=16cm}}
\caption{The ratio between Rabi frequencies $\Omega_n/\Omega_0$
experimentally measured in Ref.~\protect\cite{wine},
and well fitted by the theoretical prediction of Eq.~(\protect\ref{omn}),
is plotted as a function of the initial
vibrational number $n$ (squares), and compared with the power-law
approximation of Eq.~(\protect\ref{o35}), $(n+1)^{0.35}$ (triangles).}
\label{powerlaw}
\end{figure}

\end{document}